\documentstyle[sprocl,epsfig]{article}

\def\Journal#1#2#3#4{{#1} {\bf #2}, #3 (#4)}
\def\NCA{\em Nuovo Cimento}

\def\NPB{{\em Nucl. Phys.} B}
\def\PLB{{\em Phys. Lett.}  B}
\def\PRL{\em Phys. Rev. Lett.}
\def\PRD{{\em Phys. Rev.} D}
\def\ZPC{{\em Z. Phys.} C}

\newcommand{\be}{\begin{equation}}
\newcommand{\ee}{\end{equation}}
\newcommand{\bel}[1]{\begin{equation}\label{#1}}
\newcommand{\bea}{\begin{eqnarray}}
\newcommand{\eea}{\end{eqnarray}}
\newcommand{\beal}[1]{\begin{eqnarray}\label{#1}}

\begin{document}

\title{REGGE POLES IN QCD}

\author{A. B. KAIDALOV}

\address{Institute of Theoretical and Experimental Physics, B.Cheremushkinskaya
25, \\ Moscow 117259, RUSSIA\\E-mail: kaidalov@vitep5.itep.ru}


\maketitle\abstracts{ Basic properties of Regge poles are reviewed.
 Regge poles are considered from both t-channel and s-channel points of
 view. The main part of the review is devoted to the Regge poles in QCD. The
 method of the Wilson-loop path integral is used to calculate
 trajectories of Regge poles for  $q-\bar q$ and gluonic states. The problem of
 the Pomeron in QCD is discussed in details. It is shown how to use the
 $1/N$-expansion for classification of reggeon diagrams in QCD. The role of
 Regge cuts in reggeon theory is discussed and their importance for
 high-energy phenomenology is emphasized. Models based on the reggeon
 calculus, 1/N-expansion in QCD and string picture of interactions at large
 distances are reviewed and are applied to a broad class of phenomena in
 strong interactions. It is shown how to apply the reggeon approach to small-x
 physics in deep inelastic scattering.}

 \section{Introduction}

 Regge poles have been introduced in particle physics in the beginning of
 60-ies~\cite{Gribov,Chew} and up to present time are widely used for
 description of high-energy interactions of hadrons and nuclei. Regge
 approach establishes an important connection between high energy scattering
 and spectrum of particles and resonances. It served as a basis for
 introduction of dual and string models of hadrons. A derivation of Regge
 poles in QCD is a difficult problem closely related to the nonperturbative
 effects in QCD and the problem of confinement. In this review I shall first
 remind the main properties of Regge poles. In the section 3 I shall address
 the problem of Regge poles in QCD using the 1/N - expansion. It will be
 shown now to relate QCD diagrams to reggeon theory based on analyticity
 and unitarity. First attempts to calculate Regge
 trajectories in QCD, using the nonperturbative method of Wilson-loop path
 integral will be reviewed. A special attention will be devoted to the
 problem of the Pomeron - a Regge pole which determines an asymptotic
 behavior of high-energy diffractive processes. This problem is closely
 connected to a calculation of spectra of glueballs in QCD. An analytic formula
 for masses of glueballs will be obtained and compared to the results of recent
 lattice calculations.
 It will be shown that mixing of gluonic and light $q\bar q$-states
 is important for the Pomeron trajectory in the small t region. A role of small
 distance dynamics and results of recent perturbative calculations of the
 Pomeron will be discussed briefly. In the Section 5 the properties of Regge
 cuts will be discussed and it will be shown that these singularities play an
 important role at high energies. The last section will be devoted to the
 physics of  low-x deep inelastic scattering. This interesting kinematical
 region has been recently studied experimentally at HERA accelerator. It will
  be shown how the reggeon theory and QCD evolution effects allow to understand
  the properties of the structure function of the proton and diffractive
  dissociation of a virtual photon in a broad region of virtualities $Q^2$.\\

 \section{Regge poles and their properties}

\subsection{The reggeon concept}

The complex angular momentum method was first introduced by Regge in
 nonrelativistic quantum mechanics.\cite{Regge} In relativistic theory it
  connects a high energy behavior of scattering amplitudes with the
  singularities in the complex angular momentum plane of the partial wave
  amplitudes in the crossed (t) channel. This method is based on the general
  properties of the S-matrix - unitarity, analyticity and crossing. The
  simplest singularities are poles (Regge poles). A Regge-pole exchange is a
  natural generalization of a usual exchange of a particle with spin $J$ to
  complex values of $J$. So this method established an important connection
  between high energy scattering and the spectrum of hadrons. This is a
  t-channel point of view on Regge poles. On the other hand asymptotic
  behavior of scattering amplitudes at very high energies is closely
  related to the  multiparticle production. This is the s-channel point of
  view on reggeons.\\
      Let us consider first the t-channel point of view in more details.\\
       The binary reaction  $1+2\to 3+4$ (Fig.1) is described by the amplitude
  $T(s,t)$, which depends on invariants $s=(p_1+p_2)^2$ and $t=(p_1-p_3)^2$.
  At high energies $s\gg m^2$ and fixed momentum transfer $t\sim m^2$ an
  exchange by a particle of spin $J$ in the t-channel (Fig.2a)) leads to an
  amplitude of the form
  \bel{1}
  T(s,t)=g_{13}\cdot g_{24}\cdot (s)^J/ (M^2_J-t)
  \ee
  where $g_{ik}$ are the coupling constants and $M_J$ is the
  mass of the exchanged particle.\\
  \begin{figure}
\centering
\includegraphics[width=0.7in]{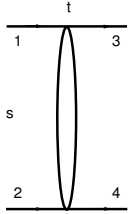}
\vskip 0.1in
\caption{A diagram for a binary reaction.
\label{fig:binary}}
\end{figure}
  For a partial wave expansion of amplitudes
\bel{2}
    f(s,cos \theta )=\frac{T(s,t)}{8\pi\sqrt{s}}=
  \frac{1}{p}\sum_{l=0}^\infty(2l+1)f_l(s)P_l(cos \theta )
\ee
  the unitarity relation leads to the constraints
\bel{3}
  \Im f_l(s)\geq \mid f_l(s)\mid^2;~~~~\mid f_l(s)\mid\leq 1
\ee
 It follows from eq.(1) that for an exchange of a particle with a spin $J\geq 2$ the
 partial wave amplitude increases with energy $\sim s^{(J-1)}$ for large $s$ and
 violates unitarity as $s\rightarrow\infty$.
\begin{figure}
\vskip0.05in
\centering
\includegraphics[width=2.0in]{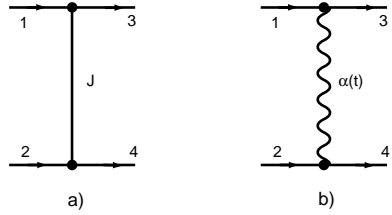}
\vskip 0.1in
\caption{a) Exchange by a particle of spin J in the t-channel.
b) Exchange by a Regge pole in the t-channel.
\label{fig:exchange}}
 \end{figure}
  This problem can be solved by introduction of Regge poles.
   It should be taken into account that the expression (1) for
 the amplitude
 is valid, strictly speaking, only close to the pole
 position $t\approx M_J^2$ and can be strongly modified away from the pole.
  Regge pole model gives an exact form of this modification and absorbs
 in itself exchanges by states of different spins (Fig2b)).
 The corresponding amplitude has the form
 \bel{4}
 T(s,t)=f_{13}(t)\cdot f_{24}(t)\cdot (s)^{\alpha (t)}\cdot \eta (\alpha (t))
 \ee
 where $\alpha (t)$ is the Regge-trajectory, which is equal to spin $J$ of
 the corresponding particle at $t=M^2$. The function
 $\eta (\alpha (t))=-(1+\sigma  exp(-i\pi\alpha (t))/\sin(\pi\alpha(t))$ is a
 signature factor
 and  $\sigma =\pm 1$ is a signature. It appears due to the fact that in
 relativistic theory it is necessary to consider separately analytic
 continuation of partial wave amplitudes in the t-channel to complex values
 of angular momenta $J$  from even $(\sigma =+)$ and odd $(\sigma =-)$ values
 of $J.$ This factor is closely related to the crossing properties of
 scattering amplitudes under interchange of $s$ to $u=(p_1-p_4)^2$. Amplitudes
 with $\sigma= +$ are even  under the interchange $s\leftrightarrow
 u ~(s\leftrightarrow -s$ for high energies), while for $\sigma=-$ amplitudes
 are antisymmetric under this operation. It should be emphasized that the
 single Regge exchange corresponds to an exchange of particles or resonances
 which are "situated" on the trajectory $\alpha (t).$  For example if
 $\alpha (t)=J,$ where $J$ is an even (odd) integer for  $\sigma =+(-)$ for
 $t=M^2_J$ and $M_J^2$ is less than the threshold for  transition to several
 hadrons ($4m_\pi^2$ for particles which can decay into two pions), then the
 Regge amplitude eq.(4) transforms into the particle  exchange amplitude
 eq.(1) with
 $g_{13}g_{24}=f_{13}(M_J^2)\cdot f_{24}(M_J^2)2/\pi\alpha' (M_J^2).$

  If $M_J$ is larger than the threshold value then $\alpha(t)$ is a complex
 function and can be written for $t\approx M_J^2$ in the form
  \bel{5}
 \alpha (t)=J+\alpha'(M_J^2)\cdot (t-M_J^2)+iIm\alpha (M_J^2)
 \ee
 In this case for $Im\alpha (M_J^2)\ll 1$ Regge pole amplitude eq.(4)
 corresponds to an exchange in the t-channel by a resonance and has the
 Breit-Wigner form
 \bel{6}
 T(s,t)=-g_{13}\cdot g_{24}(s)^J/(t-M_J^2+iM_J\Gamma _J)
 \ee
 with a width $\Gamma _J=Im\alpha (M_J^2)/M_J\alpha^{\prime}(M_J^2).$

 Thus a reggeization of particle exchanges leads to a natural resolution of
 the above mentioned problem with a violation of the unitarity,
 -Regge trajectories,  which correspond to particles with high spins can have
 $\alpha (t)\le 1$ in the physical region of high energy scattering $t<0$ and
 the corresponding amplitudes will increase with s not faster than $s^1$,
 satisfying the unitarity.  The experimental information on spectra of hadrons
 and high energy scattering  processes nicely confirms this theoretical
 expectation. The only exception is the Pomeranchuk pole (or the Pomeron),
 which determines high energy behavior of diffractive processes. I shall pay
 a special attention to properties of the Pomeron in this review.\\

 \subsection{ Bosonic and fermionic Regge poles, vacuum exchange}

 Let us consider the main properties of Regge poles.\\
 a)Factorization. Regge poles couple to external particles in a factorizable
 way, which is manifest in eq.(4).\\
 b)Regge poles have definite conserved quantum numbers like the baryon quantum
 number, parity $P$, isospin e.t.c. As it was mentioned above they have also
 a definite signature $\sigma$.

 An information on trajectories of Regge poles can be obtained for $t<0$ from
  data on two-body reactions at large $s$ and for $t>0$ from our knowledge
 of the hadronic spectrum. We have seen that a bosonic trajectory corresponds
 to particles and resonances for those values of $t$ where it passes integer
 values $(Re\alpha (t)=n)$ even for $\sigma =+$ and odd for $\sigma =-$. While
 for fermionic trajectories particles correspond to $Re\alpha (t)=\frac{n}{2}=J$
 and signature  $\sigma =(-1)^{J-\frac{1}{2}}$.\\
   There can be many trajectories with the same quantum numbers indicated
 above, which differ by a quantum number analogous to the radial quantum
 number. Such trajectories are usually called "daughter" trajectories and
 masses of corresponding resonances (with the same value of $J$) for them are
 higher than those for the leading trajectory with given quantum numbers.

 Trajectories for some well established bosonic Regge poles are shown in Fig.3.
 Note that all these trajectories have $\alpha_i(0)\leq 0.5$ for $t\leq 0$.
 One of the most interesting properties of these trajectories is their surprising
  linearity. This usually interpreted as a manifestation of strong  forces
 between quarks at large distances, which lead to color confinement.
  The linearity of Regge trajectories indicates to a string picture of the
 large distance dynamics between quarks and it was a basis of dual models
  for hadronic interactions.
\begin{figure}
\vskip0.05in
\centering
\includegraphics[width=2.5in]{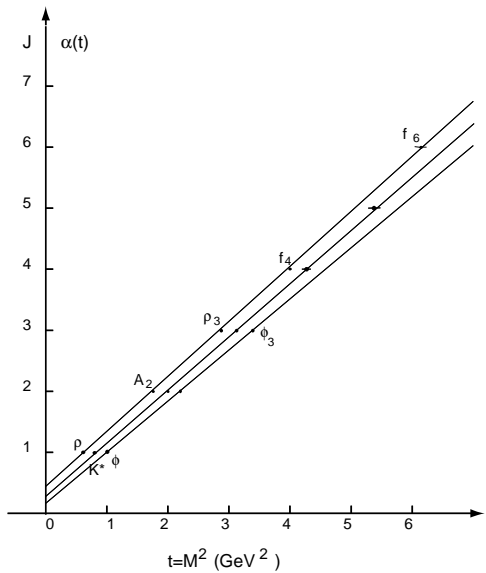}
\caption{ Trajectories for some well-known Regge poles.
\label{fig:trajectories}}
\end{figure}
 Other properties of mesonic Regge trajectories, which are evident from Fig.3
 are exchange and isospin degeneracies, - trajectories with different signatures
 and I=0 or I=1 (but with the same $\sigma P$) are degenerate with a good
 accuracy (at least in the region $t>0$). This is also in an agreement with
 dual models or approaches based on 1/N-expansion in QCD. In dual or string
 models of hadrons the daughter trajectories must be parallel to the leading
 one and displaced in the $j$-plane by integers. Experimental information on
 the daughter trajectories is rather limited but it does not contradict to these
 predictions.

 Information on mesonic Regge trajectories in the region of negative $t$,
 obtained from an analysis of binary reactions at high energies fits quite well
 the lines shown in Fig.3 obtained from the spectrum of resonances. The most
 detailed information exists for $\rho $ and  $A_2$-trajectories, which
 contribute to the reactions $\pi^-p\to \pi^0n$ and $\pi^-p\to \eta n$
 correspondingly.

 Fermionic Regge trajectories are in general analytic functions of $W=\sqrt{t}$
 and using the analyticity properties of the corresponding partial-wave
 amplitudes it is possible to show that there must be pairs of trajectories
 with different parity, which satisfy to the condition
\bel{bar}
 \alpha_+(W)=\alpha_-(-W)
\ee

Experimental data on spectrum of baryons show that baryonic trajectories
 as well as mesonic ones are nearly straight lines in variable t with the
 universal slope $\alpha'\approx 1 GeV^{-2}.$
 This universality of the slopes is natural in the string picture of baryons
 with a quark and a diquark at the ends. However, according to eq.\ref{bar}
 the fermionic trajectory, which is linear in $t=W^2$, does not change as
 $W\to -W$ and thus should coincide with its parity partner. This in its turn
 leads to parity doubling of states on this Regge trajectory.
 Experimentally  there are many baryonic states with the same spin and different
 parity, which are nearly degenerate in mass. However there are no partners
 for the lowest states ($N,\Delta $) on these trajectories.
This pattern of baryonic Regge trajectories is not yet understood
 theoretically. The relation \ref{bar} is a consequence of analyticity
 in relativistic quantum theory, but it is not realized in the existing quark
 models of baryons.

At the end of this section I shall discuss properties of the pole which has a
 special status in the Regge approach to particle physics - the Pomeranchuk
 pole or the Pomeron. This pole was introduces into theory in order account for
 diffractive processes at high energies. In the Regge pole model an amplitude
 of high energy elastic scattering has the form of eq.(4) and the total
 interaction cross section, which by the optical theorem is connected to
 $ImT(s,0),$ can be written as a sum of the Regge poles contributions
\bel{sigma}
 \sigma ^{tot}(s)=\sum_k b_k(0)(s)^{\alpha _k(0)-1}
\ee
 The poles, which are shown in Fig.3 have $\alpha _k(0)<1$ and thus their
 contributions to $\sigma ^{tot}(s)$ decrease as a $s\to \infty$. However
 experimental data show  that at $s\sim 100~ GeV^2$ total cross sections of
 hadronic interactions have a weak energy dependence and they slowly
 (logarithmically) increase with energy at higher energies. In the Regge
 pole model this can be related to the pole, which has an intercept
 $\alpha _P(0)\approx 1.$ This pole is usually called
Pomeron or the vacuum pole, because it has the quantum numbers of
 the vacuum, - positive signature, parity and G (or C) parity and isospin I=0.

 It is believed that in QCD this pole is related to gluonic exchanges in the
 t-channel. So it is usually assumed that gluonium states correspond to this
 trajectory in the region of positive $t$. We shall discuss  possible
 relation between QCD and Regge theory in more details in the next chapters.

 A value of intercept of the Pomeranchuk pole is of crucial importance for the
 Regge theory. If $\alpha_P(0)=1$, as it was assumed initially, then all the
 total interaction cross sections tend to a constant at very high energies.
 This theory has however some intrinsic difficulties  and must satisfy to many
 constraints in order to be consistent with unitarity. Besides
 experimental data indicate that $\sigma_{hN}^{(tot)}$ rise with energy.
 This logarithmic increase of total cross sections at very high energies is
 in accord with $ln^2s$ behavior consistent with the Froissart theorem.\cite{Frois}
 Thus at present the supercritical Pomeron theory with $\alpha _P(0)>1$ is
 widely used. In a model with only Regge poles taken into account an
 assumption that $\alpha_P(0)>1$ would lead to a power like increase of total
 cross sections, thus violating the Froissart bound. However in this case
 other singularities in the $j$- plane,- moving branch points should be taken
 into account and their contributions allow one to restore unitarity and to
 obtain the high energy behavior of scattering amplitudes which satisfy to the
 Froissart bound. Properties of these moving cuts are considered in the
 section 5.

 Let us note that the Pomeranchuk singularity has a positive signature,
 so it gives equal contribution to amplitudes for elastic scattering of particle
 and antiparticle. Thus it automatically satisfies to the Pomeranchuk theorem
on asymptotic equality of total interaction cross sections for particle and
antiparticle.
 A difference between these amplitudes in the Regge model is connected to
 poles with negative signature (like $\rho$ and $\omega$). It is usually assumed
 that the only singularities in the j-plane with negative signature are due
 to the known Regge poles with $\alpha _k(0)\le 0.5.$ In this case differences
of cross sections for $ a N$ and $\bar a N$-interactions decrease with energy
 as $1/\sqrt{s}$. This behavior agrees with existing experimental information.
 In principle it is possible to have a singularity with negative signature at
 $j\approx 1$. This singularity
 usually called "odderon". It appears in perturbative QCD calculations.

\subsection{s-channel picture of reggeons}

Regge poles give contributions to imaginary parts of two-body scattering amplitudes.
 For the Pomeron with $\alpha _P(0)\approx 1$ the amplitude is mostly imaginary.
Unitarity relates imaginary parts of two-body amplitudes to sums over intermediate
 states in the s-channel. So the natural question is: what are the intermediate
 states connected to reggeons? These are so called multiperipheral states the
 properties of which I shall briefly discuss now.
 \begin{figure}
\centering
\includegraphics[width=2.0in]{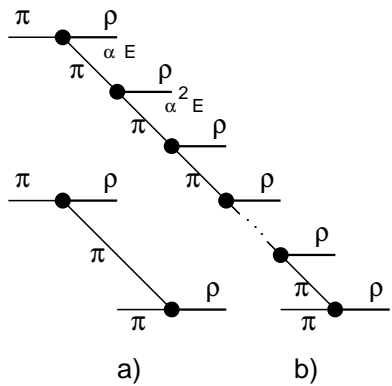}
\caption{a) Exchange by a pion in the t-channel in the reaction $\pi\pi\to\rho\rho$.
b) Diagram of multiperipheripheral production of $\rho$ mesons .
\label{fig:multiper}}
\end{figure}
Consider as an example the simplest inelastic process of $\rho $-mesons production
 in $\pi \pi $ collisions. The diagrams with pion exchange shown in Fig.4
give an important contribution to amplitudes of these processes.
 They lead to peripheral configurations - all $\rho $-mesons are produced with
 rather small momentum transfer $\sim m_\pi $.
The diagram of Fig. 4a) is large only at rather low energies when the average
 energy of the virtual pion in the lab. system ($\sim \alpha E, \alpha \leq 1$,
$ E$-is the lab. energy of the projectile) is low enough to produce a $\rho$
 meson with a target pion.  At higher energies the cross section of this
 process decreases with energy as $1/s^2$ .
At these energies it is necessary to decrease the energy of the virtual pion
 in several steps as it is shown in Fig.4b) in order to have a slow virtual
 pion at the end of the chain. If the energy decreases by a factor $\alpha $
 at each step then after n steps it will be $\alpha^nE $ and we require
 that it is $\sim m_\rho $. So on average the number of steps (related to the
 number of produced resonances)
 $\langle n\rangle \sim ln\frac{E}{m_{\rho}}/ln\frac{1}{\alpha}.$

This is an example of the diagrams of the multiperipheral model~\cite{Amati}
 of multiparticle production. Summation of these diagrams leads to the
 Regge-type behavior for an imaginary part of two-body amplitudes. So in this
 model reggeon corresponds to a sum of ladder-type diagrams shown in Fig.5.
\begin{figure}
\centering
\includegraphics[width=1.5in]{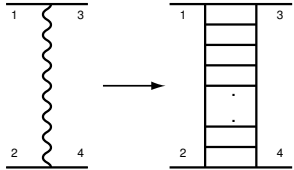}
\caption{ Reggeon as a ladder diagram.
\label{fig:ladder}}
\end{figure}
Let us consider now the multiperipheral process in the impact parameter
 $\vec b$ space. Each step has a finite size in $\vec b\sim 1/m_\pi$.
 They add independently, so there is a random walk in the $\vec b$-space
 and the average size of interaction increases with energy as
\bel{size}
\bar R^2\equiv \langle\vec b^2\rangle\approx \frac{\bar n}{m_\pi^2}\sim
 \frac{ln\frac{E}{m_{\rho} }}{m_\pi^2}
\ee
The total time of development of the fluctuation in the lab. frame is large,
 - $\tau \sim E/m^2.$

In the multiperipheral model the hadronic final states have the following
 properties.\\
a) Short range correlations in rapidity. Particles separated by several steps
 of the multiperipheral process and having substantially different energies
 [this means that difference of rapidities
 $y_1-y_2~(y=ln(E+p_{\parallel }) /m_{\perp}$) for these particles is large]
 are uncorrelated. The correlation function exponentially decreases with
 rapidity difference
\bel{cor}
c(y_1,y_2)=\frac{d\sigma }{\sigma ^{in}dy_1dy_2}
 -(\frac{d\sigma }{\sigma ^{in}dy_1})(\frac{d\sigma }{\sigma ^{in}dy_2})
 \sim exp [-\lambda (y_1-y_2)]
\ee
This property of the multiparticle final state leads to many consequences for
 inclusive cross sections. Consider for example single particle inclusive
 cross section
\bel{inc}
f^a \equiv E\frac{d^3\sigma ^a}{d^3p}=\frac{d^2\sigma
 (y_1-y_a, y_a-y_2, p^2_{\perp a})}{dy_ad^2p_{\perp a}}
\ee
  Short range correlation between particles means in this case that all inclusive
 densities $n^a\equiv f^a/\sigma ^{in}$ are independent of $y_i-y_k$ for $y_i-y_k\gg 1.$
 For example at high energies when $y_1-y_2\simeq ln (s/m^2)\gg 1$ in the
 fragmentation region of particle 1 $y_1-y_a\sim 1$ and $y_a-y_2\approx y_1-y_2\gg 1$
 the density $n^a$ becomes a function of only two variables
\bel{short}
n^a=\phi (y_1-y_a, p_{\perp a}^2)
\ee
b) The property eq.\ref{short} is equivalent to the Feynman scaling in variable
 $x_F=2p_{\parallel a}/\sqrt{s}\approx exp[-(y_1-y_a)]$
\bel{scale}
n^a=\phi (x_F, p_{\perp a}^2)
\ee
c) In the central rapidity region when both $y_1-y_a$ and $y_a-y_2$ are large
 $n^a$ becomes a function of only one variable
$(p_{\perp a}^2)$. So in this model rapidity distributions in the central
 region are flat .\\
d) The average number of produced particles $\langle n \rangle$ increases
 logarithmically with energy at large $ln (s/m^2)$.\\
e) There is a fast decrease of distributions with $p_\perp$. The form of this
 function depends on details of the model.\\
f) Multiplicity distributions of produced particles in these models have
 Poisson-type behavior.
\bel{pois}
\langle n^2\rangle -\langle n \rangle ^2\approx c\langle n \rangle
\ee
This is a consequence of short range correlations in rapidity.

The properties of inclusive distributions listed above can be quantified in
 the Regge model, using Mueller-Kancheli~\cite{Muel} diagrams .
There is a relation between a discontinuity of $3\to 3$ forward scattering
 amplitude $1\bar a 2\to 1\bar a 2$ and inclusive cross section $f^a$
  analogous to the optical theorem which relates a forward elastic scattering
 amplitude to a total cross section. For large $y_i-y_a~(s_{i\bar a}\gg m^2)$
 it is possible to use the Regge pole theory, which gives predictions on energy
 (rapidity) dependence of inclusive cross sections. For example consider the
central region of rapidities when both
 $y_1-y_a\gg 1$ and $y_a-y_2\gg 1$. In this case diagrams of double Regge
 limit (Fig.6) can be used and inclusive cross sections can be written as follows
\begin{figure}
\vskip0.05in
\centering
\includegraphics[width=0.7in]{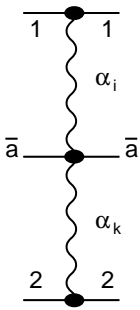}
\caption{ Regge diagram for inclusive cross section in central region.
\label{fig:central}}
\end{figure}

\bel{inccen}
 f^a=\sum_{i,k} g^i_{11}(0)g^k_{22}(0)g_{ik}^a exp[(\alpha _i(0)-1)(y_1-y_a)
 +(\alpha _k(0)-1)(y_a-y_2)]
\ee
These results can be easily obtained in the multiperipheral model.

\subsection{Diffractive production processes}

Let us consider now diffractive production of particles at high energies.
 In the Regge pole model these processes are described by the diagrams with Pomeron
exchange (Fig.7 ). It is possible to have excitation of one of the colliding
 hadrons (Fig.7a) ), - single diffraction dissociation or excitation of both
 initial particles,- double diffraction dissociation (Fig.7b)).
\begin{figure}
\vskip0.05in
\centering
\includegraphics[width=1.5in]{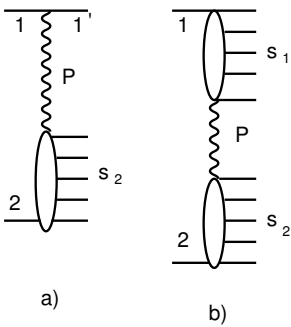}
\caption{ Diagrams for diffractive production of hadrons in the Regge pole model.
\label{fig:diffr}}
\end{figure}

For all diffractive processes there is a large rapidity gap between groups of
 produced particles. For example for single diffraction dissociation there is
a gap between the particle $1^{\prime}$ and the rest system of hadrons.
 This rapidity gap $\Delta y\approx ln (1/1-x),$ where $x$ the $x_F$ for
 hadron $1^{\prime}$ in Fig.7a). A mass of diffractively excited state at
 large $s$ can be large. The only condition for diffraction dissociation is
$s_i\ll s$. For large masses of excited states $s_2\approx (1-x)s$ and
 $\xi^{\prime}\approx \Delta y\approx ln(s/s_2).$

The cross section for inclusive single diffraction dissociation in the Regge
 pole model can be written in the following form
\bel{diffcross}
\frac{d^2\sigma }{d\xi _2dt}=\frac{(g_{11}(t))^2}{16\pi}
|G_p(\xi^{\prime},t)|^2\sigma_{P2}^{(tot)}(\xi_2,t)
\ee
where $\xi _2\equiv ln(s_2/s_0)$ and
 $G_p(\xi^{\prime},t)=\eta (\alpha _p(t))exp[(\alpha _p(t)-1)\xi^{\prime}]$
 is the Pomeron Green function. The quantity $\sigma _{P2}^{(tot)}(\xi _2,t)$
 can be considered as the Pomeron-particle total interaction cross section.\cite{Kai}
 Note that this quantity is not directly observable one and it is defined by its
 relation to the diffraction production cross section (eq.\ref{diffcross} ).
 This definition is useful however because at large $s_2$ this cross section
 has the same Regge behavior as usual cross sections
\bel{cross}
\sigma _{P2}^{(tot)}(s_2, t)=\sum_k g^k_{22}(0)r^{\alpha _k}_{PP}(t)
(\frac{s_2}{s_0})^{\alpha _k(0)-1}
\ee
where the   $r^{\alpha _k}_{PP}(t)$ is the triple-reggeon vertex (Fig.8 ),
 which describes coupling of two Pomerons to reggeon $\alpha _k$.

In this kinematical region $s\gg s_2\gg m^2$ the inclusive diffractive cross
 section is described by the triple-Regge diagrams of Fig.8 and has the form
\bel{triple}
f^1=\sum_k G_k(t)(1-x)^{\alpha _k(0)-2\alpha_P(t)}
(\frac{s}{s_0})^{\alpha _k(0)-1}
\ee
\begin{figure}
\vskip0.05in
\centering
\includegraphics[width=1.5in]{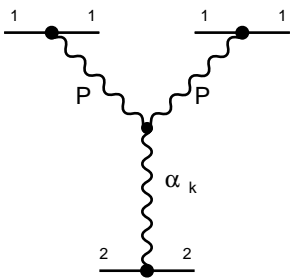}
\caption{ Triple-Regge diagram.
\label{fig:triple}}
\end{figure}
The Pomeron-proton total cross section and triple-Regge vertices $r^P_{PP},
 r^f_{PP}$ have been determined from analysis of experimental data of
 diffractive production of particles in hadronic collisions
 (see reviews~\cite{Kaidiff}).

\section{ Mesonic Regge poles in QCD }

An astonishing linearity of trajectories for Regge-poles corresponding to the
 known $q\bar q$-states indicates to an essentially nonperturbative,
  string-like dynamics. A nonperturbative method, which can be used in QCD for
 description of large distance dynamics, is the $1/N$-expansion
 (or topological expansion).\cite{Hooft,Ven}

In this approach the quantities $1/N_c$\cite{Hooft} or $1/N_{lf}$\cite{Ven}
( $N_{lf}$ is the number of light flavors) are considered as small parameters
 and amplitudes and Green functions are expanded in terms of these quantities.
 In QCD $N_c=3$ and $N_{lf}\simeq 3$ and the expansion parameter does not look
 small enough. However we shall see below that in most cases the expansion
 parameter is $1/N_c^2\sim 0.1$.

In the formal limit $N_c\to\infty$ ($N_f/N_c\to 0$) QCD has many interesting
 properties and has been intensively studied theoretically. There is a hope
 to obtain an exact solution of the theory in this limit (2-dimensional QCD
 has been solved in the limit $N_c\to\infty$).
However this approximation is rather far from reality, as resonances in this
 limit are infinitely narrow ($\Gamma \sim 1/N_c$). The case when the ratio
 $N_f/N_c\sim 1$ is fixed and the expansion in $1/N_f$ (or $1/N_c$) is carried
 out~\cite{Ven}  seems more realistic.

This approach is called sometimes by the topological expansion, because the
 given term of this expansion corresponds to an infinite set of Feynman
 diagrams with definite topology. It should be emphasized that $1/N$-expansion
 should be applied to Green-functions or amplitudes for white states.

The first term of the expansion corresponds to the planar diagrams of the type
 shown in Fig.9  for the binary reaction. These diagrams always have as border
 lines the valence quarks of the colliding hadrons. At high energies they should
 correspond to exchanges by secondary Regge poles
 $\alpha _R (\rho , A_2,\omega ,...)$ "made of" light quarks.
 The s-channel cutting of the planar diagram of Fig.9a)  is shown in Fig.9b).
Here and in the following we do not show internal lines of gluons and quark loops.
 This diagram corresponds to a multiparticle production, which has the same
 properties as in the multiperipheral model.
\begin{figure}
\vskip0.05in
\centering
\includegraphics[width=2.0in]{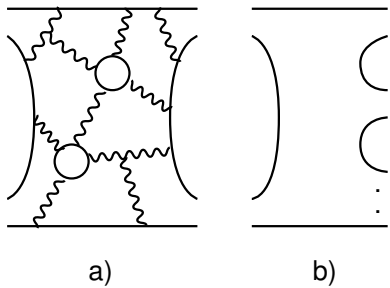}
\caption{ a) Planar diagram for binary reactions. Full lines denote quarks,
wavy lines - gluons. b) Same for $12\to X$. Internal lines of gluons and quarks
are not shown.
\label{fig:planar}}
\end{figure}
The topological classification of diagrams in QCD leads to many relations between
 parameters of the reggeon theory, hadronic masses, widths of resonances and
 total cross sections (for a review see~\cite{Ksoft}) . All these relations are
 in a good agreement with experiment.

A contribution of the planar diagrams to the total cross section decreases with
 energy as $1/s^{(1-\alpha _R(0))}\approx 1/\sqrt{s}.$
This decrease is connected to the fact that quarks have spin 1/2 and in the
 lowest order of perturbation theory an exchange by two quarks in the t-channel
 leads to the behavior $\sigma \sim 1/s$, which corresponds to the intercept
 $\alpha _R(0)=0.$ Interaction between quarks should lead to an increase of
 the intercept to the observed value $\alpha _R(0)\approx 0.5$.

Is it possible to calculate Regge-trajectories from QCD? Even for planar
 diagrams this is a difficult problem. It was considered in paper~\cite{Dub}
 using the method of Wilson-loop path integral.\cite{Sim} It was shown that under
 a reasonable assumption about large distances dynamics: minimal area law for
 Wilson loop at large distances, confirmed by numerous lattice data,
 $\langle W\rangle \sim exp (-\sigma  S_{min})$,
 it is possible to calculate spectrum of  $q\bar q$-states.
 In these calculations virtual $q\bar q$-pairs and spin effects were neglected.
 It was shown that the mass spectrum can be determined from the following effective
 Hamiltonian $H_0$
 $$  H_0=\frac{p_r^2}{\mu(t)}+
\mu(t)+\frac{L(L+1)}{r^2[\mu+2\int^1_0(\beta-\frac12)^2\nu d\beta]}+$$
 \bel{Ham}
 +\int^1_0\frac{\sigma^2_{adj}d\beta}{2\nu(\beta,t)} r^2 +\frac12
 \int^1_0 \nu(\beta, t) d\beta
 \ee
 Here $\mu(t)$ and $\nu(\beta,t)$ are positive auxiliary functions which are
 to be found from the  extremum condition.\cite{Dub}  Their extremal values
 are equal to the  effective quark energy $\langle \mu\rangle $ and energy
 density of the adjoint string $\langle \nu\rangle $.

 The resulting spectrum of $H_0$ for light quarks with a good accuracy is
  described by a very simple formula
 \bel{mass}
 \frac{M^2}{2\pi\sigma}=L+2n_r+c_1
 \ee
\begin{figure}
\vskip0.05in
\centering
\includegraphics[width=2.5in]{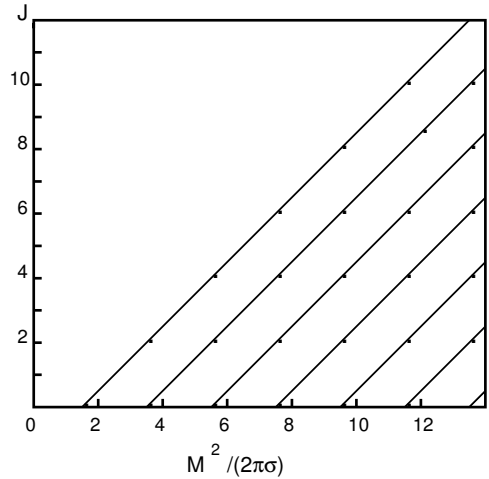}
\caption{Calculated spectrum of $q-\bar q$ states and corresponding Regge-trajectories.
\label{fig:spectrum}}
\end{figure}

 This spectrum is shown in Fig.10 and corresponds to an infinite set of
 linear Regge trajectories similar to the one of dual and string models.\\
 In order to make realistic calculations of masses of hadrons which can be
 compared with experiment it is necessary to take into account perturbative
 interactions at small distances, spin effects and quark loops. For light
 quarks spin effects are non trivial as the spontaneous violation of the
 chiral symmetry should be properly taken into account.

 \section{Glueballs and the Pomeron in QCD }

 It was mentioned above that the Pomeron in QCD is usually related to gluonic
  exchanges in the $t$-channel.\cite{Low}. This is connected with the fact that
  gluons naturally lead to the vacuum quantum numbers and the simplest
  perturbation theory diagram for scattering amplitude with an exchange by
  2 gluons leads to a cross section which does not depend on energy
  (due to spin of gluon $Sg=1$). Thus in this approximation $\alpha _P(0)=1.$
  In perturbation theory an interaction between gluons leads to an increase of
  the Pomeron intercept.\cite{BFKL}

From the point of view of 1/N-expansion the Pomeron is related to the cylinder-type
 diagrams, shown in Fig.11 with gluonic states (mixed with $q\bar q$-pairs)
 on a surface. In this approach the Pomeron is related to glueballs.
\begin{figure}
\vskip0.05in
\centering
\includegraphics[width=1.5in]{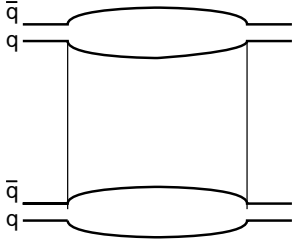}
\caption{ Cylinder-type diagrams for the Pomeron.
\label{fig:cylinder}}
\end{figure}

  Glueballs are among the most  intriguing objects both in experiment
  and theory. While experimental situation is not yet
 settled, lattice simulations~\cite{Lat1,Lat2} yield an overall consistent picture
 of lowest ($<4~ GeV$) mass spectrum.
 The mass scale and level ordering of the resulting glueball
 spectra strongly differ from those of meson spectra, yielding a
 unique information about the nonperturbative  structure of the
 gluonic vacuum.

The problem of spectra of glueballs and its relation to the Pomeron was considered
 in papers~\cite{Simg,Simn}, using the method of Wilson-loop path integral
  discussed above for the case of $q\bar q$-Regge poles.

In the approximation when the spin effects and quark loops are neglected the
spectrum of two-gluon glueballs is determined by the same Hamiltonian $H_0$
given by eq.\ref{Ham} with the only difference that the string tension
($\sigma _{fund}$) for the $q\bar q$ system is changed to $\sigma_{adj}$.
Thus the mass spectrum for glueballs is given be eq.\ref{mass} with the change
$\sigma \rightarrow \sigma _{adj}$.

The value of $\sigma_{adj}$ in (20) can be found from the string
tension $\sigma_{fund}$ of $q\bar q$ system,
multiplying it  by $\frac{9}{4}$, as it follows from Casimir scaling
observed on the lattices. Taking  experimental Regge slope for
mesons $\alpha'=0.89~ GeV^{-2}$ one obtains $\sigma_{fund}=0.18~ GeV^2$
and $\sigma_{adj}\approx 0.40~ GeV^2$.

In order to compare our results with the corresponding lattice calculations~\cite
{Lat1,Lat2} it is convenient to consider the quantity
$\bar M/\sqrt{\sigma_f}$, which is not sensitive to the choice of
string tension $\sigma$.
 We also introduce the spin averaged mass $\bar M$ which for $L=0, n_r=0$ states
  is defined as $\bar M=\frac13 (M(0^{++})+2M(2^{++}))$, and in a similar way for
higher states. This definition takes into account the structure of
spin-splitting terms, so that $\bar M$ can be
compared with the eigenvalues $M_0$ of spinless Hamiltonian eq.\ref{Ham}.

The comparison of our predictions for spin averaged masses of the
lowest glueball states with corresponding lattice results is given in
Table 1.  For the average mass with  $L=2, n_r=0$  lattice
results are limited to the state  $3^{++}$. An agreement is perfect especially
for the mass of the lowest state which is calculated on lattices with highest
accuracy.

 \newpage
\begin{center}

{\bf Table 1}\\
 Spin averaged glueball masses $M_G/\sqrt{\sigma_f}$

\vspace{0.5cm}

 \begin{tabular}{|l|l|l|l|l|} \hline
\multicolumn{2}{|c|}{ Quantum}& This& \multicolumn{2}{c|}{Lattice data}\\
\cline{4-5}
\multicolumn{2}{|c|}{ numbers}&work&paper~\cite{Lat2}&paper~\cite{Lat1}\\hline
 &$l=0,n_r=0$&4.68&4.66$\pm$0.14&4.55$\pm$0.23\\   \cline{2-5}
 2 gluon&$l=1,n_r=0$&6.0 &6.36$\pm$0.6 &6.1 $\pm$0.5 \\\cline{2-5}
 states&$l=0,n_r=1$&7.0 &6.68$\pm$0.6 &6.45$\pm$0.5 \\\cline{2-5}
 &$l=2,n_r=0$&7.0 &9.0 $\pm$0.7(3$^{++}$) &7.7 $\pm 0.4(3^{++})$ \\
 \cline{2-5}
 &$l=1,n_r=1$&8.0 &  &8.14 $\pm 0.4(2^{*-+})$ \\ \hline
 3 gluon&K=0&7.61&&8.19$\pm$0.48\\
 state&&&&\\\hline
   \end{tabular}
\vspace{1.5cm}

{\bf Table 2}\\

Comparison of predicted glueball masses with lattice data

\vspace{0.5cm}

 \begin{tabular}{|l|l|l|l|l|l|r|} \hline
$J^{PC}$& M(GeV)&  \multicolumn{2}{c|}{Lattice
 data}
 &\multicolumn{2}{c|}{$M[G]/M[0^{-+}]$}
 &Difference\\ \cline{3-4}   \cline{5-6}
  &This   work& paper~\cite{Lat1}&paper~\cite{Lat2}
  &This   work&paper~\cite{Lat1}&\\\hline

$0^{++}$&1.58&1.73$\pm$0.13&1.74$\pm$0.05
&0.62&0.67(2)&-7\%\\
$0^{++*}$&2.71&2.67$\pm$0.31&3.14$\pm$0.10
&1.06&1.03(7)&3\%\\
$2^{++}$&2.59&2.40$\pm$0.15&2.47$\pm$0.08
&1.01&0.92(1)&9\%\\
$2^{++*}$&3.73&3.29$\pm$0.16&3.21$\pm$0.35
&&&\\
$0^{-+}$&2.56&2.59$\pm$0.17&2.37$\pm$0.27
&&&\\
$0^{-+*}$&3.77& 3.64$\pm$0.24&
&1.47&1.40(2)&5\%         \\
$2^{-+}$&3.03&3.1$\pm$0.18&3.37$\pm$0.31
&1.18&1.20(1)&-1\%\\
$2^{-+*}$&4.15& 3.89$\pm$0.23&
&1.62&1.50(2)&8\%         \\
$3^{++}$&3.58&3.69$\pm$0.22&4.3$\pm$0.34
&1.40&1.42(2)&-2\%\\
$1^{--}$&3.49& 3.85$\pm$0.24&
&1.36&1.49(2)&-8\%         \\
$2^{--}$&3.71& 3.93$\pm$0.23&
&1.45&1.52(2)&-1\%         \\
$3^{--}$&4.03& 4.13$\pm$0.29&
&1.57&1.59(4)&         \\\hline
  \end{tabular}
   \end{center}
\vspace{0.5cm}
   The spin splittings for glueball masses were calculated in paper~\cite{Simn}
   assuming that spin effects can be treated as small perturbations. A largest
   correction is obtained for the lowest state from the spin-spin interaction.
   The results are compared with lattice calculations in the Table 2 (for
   $\alpha_s=0.3$).

Let us consider now the Pomeron Regge trajectory in this approach
 in more details, taking into account both  nonperturbative and perturbative
contributions to the Pomeron dynamics.

The large distance, nonperturbative contribution gives according to Eq.(20)
for the leading glueball trajectory ($n_r=0$ )
\bel{gluepom}
\alpha_P(t)=-c_1+\alpha'_P t +2
\ee
with $\alpha'_P=\frac{1}{2\pi\sigma_a}$.

In eq.\ref{gluepom} spins of "constituent"
gluons are taken into account, but a small nonperturbative spin-spin
interactions were neglected.

For the intercept of this trajectory we obtain
$\alpha_P(0)\approx 0.5$, which is substantially below the value
found from analysis of high-energy interactions
$\alpha_P(0)=1.08\div 1.2$. I would like to emphasize that contrary to interactions
related to emission of real particles in the s-channel (for example emission of
gluons in the perturbative ladder-type diagrams) the confining interaction considered
above leads to a decrease of an intercept of a Regge trajectory compared to the
Born approximation.

 The most important nonperturbative source, which can lead to an increase of the
Pomeron intercept is the quark-gluon mixing or account of quark-loops in the gluon
 "medium". In the 1/N -expansion this effect is proportional to $N_f/N_c$, where
$N_f$ is the number of light flavors.
In the leading approximation of the $1/N_c$-expansion there are 3
Regge-trajectories with vacuum quantum numbers,
- $q\bar q$-planar trajectories ($\alpha_f$ made of
$u\bar u$ and $d\bar d$ quarks, $\alpha_{f'}$ made of $s\bar s$-quarks) and
pure gluonic trajectory - $\alpha_G$. The transitions between quarks and
gluons $\sim\frac{1}{N_c}$ will lead to a mixing of these trajectories.
Note that for a realistic case of $G,f$ and $f'$-trajectories (eq.\ref{gluepom}
and Fig.3) all 3-trajectories before mixing are close to each other in the small
$t$ region. In this region mixing between trajectories is essential even for small
coupling matrix $g_{ik}(t).$ Lacking calculation of these effects in QCD they
were considered in the paper~\cite{Simn} in a semi--phenomenological manner.

 Denoting by $\bar \alpha_i$ the bare  $f,f'$ and $G$--trajectories
 and introducing the mixing matrix $g_{ik}(t)(i,k=1,2,3)$ we obtain
 the following equation for determination of resulting trajectories
 after mixing.\cite{Simn}
 \bel{mix}
   j^3-j^2\sum\bar \alpha_i+j(\sum_{i\neq k}\bar
 \alpha_i\bar \alpha_k-g^2_{ik})-
 \bar\alpha_1\bar\alpha_2\bar\alpha_3+
 \sum_{i\neq k\neq l}\bar \alpha_lg^2_{ik}-2g_{12}g_{13}g_{23}=0
 \ee
  For realistic values of $g_{ik}(t)$ (for details see~\cite{Simn}) the
  Pomeron intercept is shifted to the values $\alpha_P(0)\approx 1$. For
  $t>1 GeV^2$ the Pomeron trajectory  is very close to the planar
  $f$--trajectory, while the second and third vacuum trajectories --
  to $\alpha_{f'}$ and $\alpha_G$ correspondingly. In the region of large $t>0$
  the effects of mixing are small.

  It is interesting that with an account of the quark-gluon mixing the
 intercept of the Pomeron trajectory is close to the value $j=1$ corresponding
 to an exchange by 2 noninteracting gluons.

    Up to now I have considered mostly nonperturbative, large distance dynamics
 of the Pomeron. A small distance dynamics of the Pomeron has been studied in
 many papers using the QCD perturbation theory (see for example reviews~\cite{revBFKL}).
 in the leading log approximation the Pomeron corresponds to a sum of the
 ladder-type diagrams with  exchange of reggeized gluons in the t-channel
 (BFKL~\cite{BFKL} Pomeron). In this approximation the intercept of the Pomeron
 is equal to
\bel{LLA}
\Delta=\alpha_P(0)-1=\alpha_s\frac{12}{\pi}ln 2
\ee
It has been found recently~\cite{Fadin} that $\alpha_s$
corrections substantially decrease $\Delta$ compared to  LLA
result. The intercept of the Pomeron  depends on the renormalization
scheme and scale for $\alpha_s$. In the "physical" (BLM) scheme values of
 $\Delta$ are in the region $0.15\div 0.17$ .\cite{Brodsky} Unfortunately it
 is very difficult to calculate higher order corrections in PQCD.

  Sometimes
 the BFKL Pomeron is called "hard" Pomeron contrary to the "soft" one. However
 the equation for the Pomeron singularity contains both nonperturbative
 effects discussed above and perturbative dynamics. Thus the resulting
 "physical" pole is a state due to both "soft" and "hard" interactions.

\section{ Regge cuts. High-energy hadronic interactions}

\subsection{Multi-Pomeron exchanges}

Regge poles are not the only singularities in the complex angular momentum plane.
 Exchange by several reggeons in the $t$-channel shown for the Pomeron case in
  Fig.12 leads to moving branch points (or Regge cuts) in the j-plane.
   The positions of the branch points for t=0 can be expressed in terms of
  intercept of the Pomeranchuk pole
  \bel{cut}
\alpha _{np}(0)-1=n(\alpha _p(0)-1)=n\Delta
\ee
Contributions of these singularities to scattering amplitudes
$T_n(s,0)\sim s^{1+n\Delta }$ are especially important at high energies for
$\Delta >0$. The whole series of n-Pomeron exchanges should be summed.
\begin{figure}
\vskip0.05in
\centering
\includegraphics[width=1.5in]{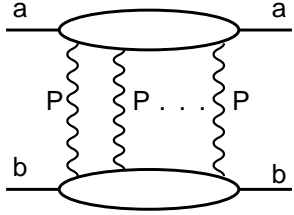}
\caption{ Diagrams with an exchange by several Pomerons in the t=channel.
\label{fig:multipom}}
\end{figure}
An account of these multi-Pomeron exchanges in the $t$-channel leads to
unitarization of scattering amplitudes.

In the framework of 1/N-expansion the n-Pomeron exchange amplitudes are due to
 topologies with (n-1) handles and are of the order $(1/N^2)^n.$

The Gribov reggeon diagrams technique~\cite{regtech} allows one to calculate
contributions of Regge cuts to scattering amplitudes. I shall illustrate this
using as an example the two-Pomeron exchange contribution to the elastic scattering
amplitude.
\begin{figure}
\vskip0.05in
\centering
\includegraphics[width=1.2in]{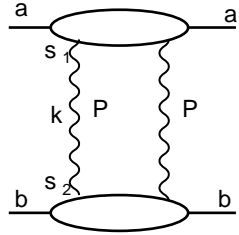}
\caption{ The diagram of two-Pomeron exchange for elastic scattering amplitude.
\label{fig:twopom}}
\end{figure}
The Pomeron-particle scattering amplitudes, which enter into the diagram of Fig.13,
 have usual analyticity properties in variables $s_i$ (poles and cuts) and changing
  the integration in the diagram from $d^4k$ to
 $ds_1ds_2d^2k_\perp/2s$ one can write the diagram of two-Pomeron exchange in the
 form
 $$T_{2P}(s,t)=\frac{i}{2!}\int\frac{d^2k_\perp}{\pi}\eta _p(k^2_\perp)\eta _p((\vec k_\perp-\vec q)^2)$$
\bel{twoP} (\frac{s}{s_0})^{\alpha_P(k_\perp ^2)+\alpha_P((\vec k_\perp-\vec q)^2)-2}\times N_a(\vec k_\perp, \vec q)
 N_b(\vec k_\perp, \vec q)
\ee
The amplitudes $T_{aP,bP}$ decrease faster than $1/s_i$ at large $s_i$ and
 the contours of integrations on $s_i$ can be transformed in such a way that only
discontinuities of these amplitudes at the right hand cut enter into $N_a, N_b$.
 The Pomeron-particle scattering amplitudes satisfy to the unitarity condition,
 and the discontinuities can be expressed as a sum over
contributions of intermediate states on mass shell.
Thus the two-Pomeron exchange diagram of Fig.13 can be expressed as the sum of
diagrams, shown in Fig.14. The diagram of Fig. 14a) corresponds to the P pole
contribution to the elastic scattering amplitude, while the diagrams of
Fig. 14 b)-d) are P pole contributions to inelastic diffraction.
\begin{figure}
\vskip0.05in
\centering
\includegraphics[width=3.0in]{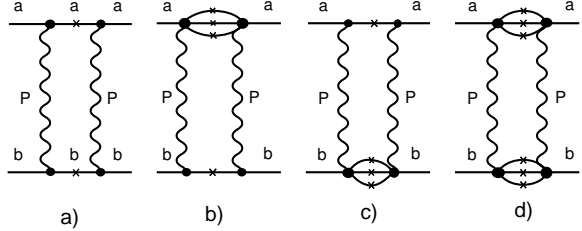}
\caption{ Rescattering diagrams. (Crosses on lines indicate that they are on
mass shell).
\label{fig:rescat}}
\end{figure}
The diagrams with n-Pomeron exchange in the t-channel can be treated in the same
 way. The sum of the elastic rescatterings leads to the eikonal
 formula for the amplitude of ab-scattering in the impact parameter space
\bel{eikonal}
 f_{ab}(s,b)=\frac{i}{2}(1-exp[2i\delta _P(s,b)])
\ee
where $\delta _P(s,b)=\int \frac{d^2q}{2\pi} exp(i\vec q\cdot \vec b)T_P(s,q^2_\perp)$.
In general $f(s,b)$ can be written in the form
\bel{rescat}
 f(s,b)=\frac{1}{2i}\sum_{n=1}^\infty\frac{(-v_p)^n}{n!}\gamma _n
\ee
where $v_P(s,b)=-2i\delta _P(s,b).$ In the eikonal approximation all
$\gamma _n=1.$ All diagrams in Fig.14 have the same sign. This leads to the
 constraint $\gamma _2\equiv C>1,$ which means that $|T_{2P}|$ for the contribution
 of the PP-cut to the amplitude is always larger than the eikonal value .
 A simplest generalization of the eikonal model, which takes
 into account also inelastic diffractive intermediate states is so called
 "quasieikonal" model, where $\gamma _n=C^{n-1}$ and the amplitude $f(s,b)$ has
 the form
\bel{quasi}
 f(s,b)=\frac{i}{2C}(1-exp(-Cv_P))
\ee
The function $v_P(s,b)\sim s^\Delta $  and it becomes large at very high energies.
 In this limit the scattering amplitude for elastic scattering $f(s,b)\to i/2$ in
 the eikonal model (scattering on a black disk) and $f(s,b)\to 1/2C$ in the
 quasieikonal model (scattering on a grey disc).
This property of the quasieikonal model is closely related to the fact that one
of the eigen states for the diffractive matrix $\hat v_P$ has in this model a
zero eigenvalue.
This is a crude approximation, which takes into account a big difference in the
interaction cross section of hadrons with different transverse sizes.
Configurations of quarks inside hadrons with small transverse size $r$ have total
 interaction cross sections $\sim r^2$, because hadrons in QCD interact as color
 dipoles.
There is a distribution of quarks and gluons inside colliding hadrons with different
 values of $r$ so one can expect that there will be a slow approach to the black
 disk limit for elastic scattering amplitude as $s\to \infty$.
In this limit the effective radius of interaction increases as ln s.
 Thus total interaction cross sections for the supercritical Pomeron theory
 have Froissart type behavior $\sigma ^{(tot)}\sim ln^2(s)$  as $s\to \infty$.

\subsection{ AGK-cutting rules}

Now we shall discuss the s-channel unitarity for contributions of Pomeron cuts.
The connection between different s-channel intermediate states and contribution
of cuts to the imaginary part of elastic scattering amplitude is given by
Abramovsky, Gribov, Kancheli (AGK)-cutting rules.~\cite{AGK} I shall illustrate
them using as an example two-Pomeron exchange amplitude (Fig.13).
Its contribution to the imaginary part of the forward elastic ab-scattering
amplitude is negative and can be denoted as $T_{2P}(s,b)=-A(s)$ and to the
$\sigma ^{(tot)}$ as $-\sigma _2.$

There are three different classes of diagrams, which can be obtained by the s-channel
 cuttings of the diagram of Fig.15. They contribute to different classes of physical
  processes.\\
\begin{figure}
\vskip0.05in
\centering
\includegraphics[width=3.0in]{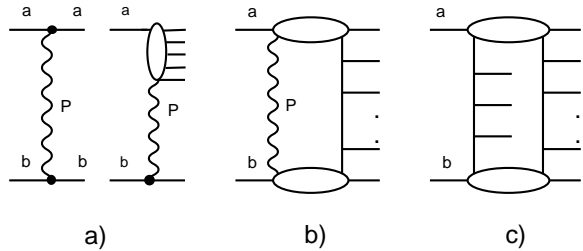}
\caption{ Different contributions to the s-channel cutting of the two-Pomeron
exchange diagram.
\label{fig:AGK}}
\end{figure}
a) The cutting between the Pomerons leads to diffractive processes (both elastic
 and inelastic), shown in Fig.15 a). Their contributions to the $ImT_{ab}(s,0)$
  and $\sigma ^{(tot)}$ according to AGK-rules are A(s) and $\sigma _2$ ñcorrespondingly.\\
b) Cutting through one of Pomerons (Fig.15 b)) leads to absorptive corrections
for the multiperipheral type inelastic production of particles, corresponding to
 the s-channel content of the Pomeron. AGK rules give for this contribution-4A(s)
and $-4\sigma _2$  correspondingly.\\
c) Cutting through both of Pomerons (Fig.15 c)) leads to a new process,- production
 of two multiperipheral chains. According to AGK-rules this contributes 2A(s) to
 the imaginary part of the forward scattering amplitude and $2\sigma _2$ to the
 total cross section.

The summary contribution of all these process to the total cross section is
$(1-4+2)\sigma _2=-\sigma _2.$

The AGK rules are formulated~\cite{AGK} for arbitrary diagrams with exchange of $n$
reggeons in the t-channel. They allow to classify all multiparticle configurations
 and calculate their weights in terms of contributions to forward elastic scattering
 amplitude (or the corresponding amplitudes in the impact parameter space). For
 example in the eikonal or quasieikonal models all cross sections (diffractive
 and inelastic with different number of multiperipheral chains) can be expressed
  in terms of a single Pomeron exchange contribution to an elastic scattering
 amplitude.

Let us consider some consequences of these rules for particle production at very high
 energies. \\
1) For a sum of all n-Pomeron exchange diagrams of the eikonal-type (without
interaction between Pomerons) there is a cancellation of their contributions to
the single particle inclusive spectra in the central rapidity region for
$n\ge 2 $ .\cite{AGK}
So only the pole diagram of Fig.6  contributes and inclusive spectra increase
with energy as $f^a\sim (s/s_0)^{\Delta} .$ This means in particular that a study
 of an energy dependence of inclusive spectra in the central rapidity region gives
  more reliable information on the value of $\Delta ,$ than $\sigma ^{(tot)},$
  where Pomeron cuts strongly modify energy dependence
compared to the pole diagram. Inclusive charged particles
density at $y=0$ for $pp(\bar pp)$-collisions has a fast increase with energy
and can be described in eikonal-type models with the intercept of the
Pomeron in the range $\Delta =0.12-0.14$. The calculation of the total and elastic scattering
 cross sections in the same model with these values of $\Delta $ also leads to a good
  description of experiment.\cite{QGSr,DTUr} It should be noted that the value of
  $\Delta $ becomes larger $(\Delta \approx 0.2)$ if the interaction between Pomerons
  is taken into account~\cite{Ponomar} (see below).\\
2) Pomeron cuts lead to long range correlations in rapidity. Existence of such
 correlations (for example long range correlations between number of particles
 produced in the forward and backward hemispheres in the c.m. system) is now
 firmly established in particle production at high energies.\\
3) Existence of poliperipheral contributions with several multiperipheral
 chains leads to broad multiplicity distributions of produced hadrons
 with the dispersion $D\sim \langle n\rangle.$ The s-channel cutting
 of a single Pomeron leads to a Poisson-type distribution with
 $\langle n\rangle =\langle n \rangle_P,$ for cutting
 of two Pomerons (Fig.14 c)) the distribution is of the same type but with
 $\langle n\rangle \approx 2\langle n \rangle_p$ and so on.
The summary contribution for cuttings of all n-Pomeron exchange diagrams is broad
 and has a complicated form. At not too high energies contributions from different
 n strongly overlap and multiplicity distributions satisfy to a good accuracy
 KNO-scaling $(\langle n \rangle \sigma _n/\sigma ^{(in)}=f(n/\langle n \rangle)).$
However as energy increases KNO scaling is violated. The models based on the reggeon
 diagrams technique and AGK-cutting rules~\cite{QGSr,DTUr} give a good
 quantitative description of multiplicity distributions .\\

The theoretical models mentioned above (the Dual Parton Model~\cite{DTUr}
 and the Quark Gluon Strings Model~\cite{QGSr} use besides the reggeon theory
  also 1/N-expansion for interpretation of different reggeon diagrams in QCD.
  This leads to a very predictive approach to multiparticle production
  in hadron -hadron, hadron-nucleus and nucleus-nucleus collisions.
With a small number of parameters these models allow one to describe total and
elastic cross sections, diffraction dissociation, multiplicity distributions,
inclusive cross sections for different types of hadrons e.t.c. in a broad region
of energies.\cite{QGSr,DTUr}. In the following we shall apply the same approach to DIS
 in the small $x$ region.\\

\subsection {Interactions between Pomerons }

In the eikonal-type models discussed above the diffraction dissociation to the
states with not too large masses has been taken into account. The diffractive
production of states with large masses is related, as we know from the discussion
 of diffractive processes, to the diagrams of the type shown in Fig.8 with
 interaction between Pomerons.  Neglect in the first approximation by these
 interactions is justified by a smallness of triple-Pomeron and 4-Pomeron
 interaction vertices, found from analysis of diffractive processes.\cite{Kaidiff}
However at very high energies it is necessary to include all these diagrams in
order to have a selfconsistent description of high energy hadronic interactions,
 including large mass diffractive production of particles. It was demonstrated
 in paper~\cite{Ponomar} that inclusion of these diagrams leads in most of the
 cases to predictions which are very close to the results of eikonal type models,
  however the value of the "bare" Pomeron intercept increases up to the value
$\alpha _P(0)\approx 1.2$.

\section {Small-x physics}

This section is devoted to a problem of small-x physics in deep inelastic
 scattering (DIS) .
 This problem became especially actual due to recent experimental
investigation of this region at HERA accelerator.

In DIS it is possible to study different asymptotic limits. For a virtuality
of the photon $Q^2\rightarrow \infty$ and $x=Q^2/(W^2+Q^2)\sim 1$ the usual QCD
evolution equations can be applied and $Q^2$ dependence of the structure functions
can be predicted
if an initial condition for structure functions at $Q^2=Q_0^2$ is formulated.
On the other hand if $Q^2$ is fixed and $x\rightarrow 0$ (or
$\ln(1/x)\rightarrow \infty$) the asymptotic Regge limit is relevant. The most
interesting question is what is the behavior of DIS in the region
where both $\ln(1/x)$ and $\ln(Q^2)$ are large? This is a transition
region between perturbative and nonperturbative dynamics in QCD and
its study can give an important information on the properties of
confinement and its relation to the QCD perturbation theory. The
asymptotic Regge limit in DIS can be related to high-energy limit
of hadronic interactions and  is described in terms of the
Pomeranchuk singularity.

 Experiments at HERA have found two extremely important properties of small-$x$
 physics~: a fast increase of parton densities as $x$ decreases~\cite{H1F2,ZEUSF2}
 and the existence of substantial diffractive production in deep inelastic scattering
 (DIS).\cite{ZEUSd,H1d}

   A fast increase of $\sigma^{(tot)}_{\gamma^*p}$ as $W^2\equiv s$ increases
 at large $Q^2$ observed experimentally~\cite{H1F2,ZEUSF2} raises a question: whether
there are two different Pomerons -- "soft" and "hard"?   From the
discussion of the Pomeron in QCD above it follows that there are no
theoretical reasons for such a situation and most probably
the rightmost pole in the $j$--plane is generated by both
 "soft" and "hard" dynamics. I shall assume that there is one ("physical")
 Pomeron pole with the same $\alpha_P(0)$ as it was determined from
 high--energy hadronic interactions with an account of many--Pomeron cuts.
 On the other hand the effective intercept, which depends on relative contribution of
multi--Pomeron diagrams, can be different in different processes.

 In paper~\cite{9r} it was suggested that the increase of the effective
 intercept of the Pomeron, $\alpha_{eff} = 1 + \Delta_{eff}$, as $Q^2$
 increases from zero to several GeV$^2$ is mostly due to a decrease of
 shadowing effects with increasing $Q^2$. A parametrization of the $Q^2$
 dependence of $\Delta_{eff}$ such that
 $\Delta_{eff} \approx 0.1$ for $Q^2 \approx 0$ (as in soft hadronic
 interactions) and $\Delta_{eff} \approx 0.2$ (bare Pomeron intercept) for
 $Q^2$ of the order of a few GeV$^2$, gives a good description of all
 existing data on $\gamma^*p$ total cross-sections in the region of
 $Q^2 \leq  5 \div 10$~GeV$^2$.\cite{9r,10r} At larger $Q^2$ effects
 due to QCD evolution become important. Using the above parametrization as
 the initial condition in the QCD evolution equation, it is possible to
 describe the data in the whole region of $Q^2$ studied at HERA.\cite{9r,11r}

 In the reggeon approach discussed above there are good reasons to believe that
 the fast
 increase of the $\sigma_{\gamma^*p}$ with energy in the HERA energy range
 will change to a milder increase at much higher energies. This
 is due to multi--Pomeron effects, which are related to shadowing in highly dense
 systems of partons - with eventual ``saturation'' of densities. This problem
 has a long history (for reviews see~\cite{revBFKL,Kr}) and has been extensively discussed
 in recent years.\cite{2r} It is closely connected to the problem of the
 dynamics of very high-energy heavy ion collisions.\cite{3r}

This problem was investigated recently in our paper~\cite{CFKS}, where reggeon
approach was applied to the processes of diffractive $\gamma^* p$ interaction.
 It was emphasized in the previous section that in the reggeon calculus~\cite{regtech}
  the amount of rescatterings is closely related to diffractive
 production. AGK-cutting rules~\cite{AGK} allow to calculate the cross-section
 of inelastic diffraction if contributions of
 multi-Pomeron exchanges to the elastic scattering amplitude are known.
 Thus, it is very important for self-consistency of theoretical models to
 describe not only total cross sections, but, simultaneously, inelastic
 diffraction. In particular in the reggeon calculus the variation of $\Delta_{eff}$
 with $Q^2$ is related to the corresponding variation of the ratio
 of diffractive to total cross sections.
In the paper~\cite{CFKS} an explicit model for the contribution of
 rescatterings was constructed which leads to the pattern of energy behavior
 of $\sigma_{\gamma^*p}^{(tot)} (W^2,Q^2)$ for different
 $Q^2$ described above. Moreover, it allows to describe simultaneously
 diffraction production by real and virtual photons. In this model it is
possible to study quantitatively a regime of "saturation" of parton
densities.

Let us discuss briefly the qualitative picture of diffractive dissociation
of a highly virtual photon at high energies. It is convenient to
discuss this process in the lab. frame, where the quark-gluon fluctuations
of a photon live a long time $\sim 1/x$ (Ioffe time~\cite{Ioffe}). A virtual
photon fluctuates first to $q\bar q$ pair. There are two different types of
configurations of such pair, depending on transverse distance between
quarks (or $k_{\perp})$.\\
a) Small size configurations with $k^2_{\perp}\sim Q^2$. These small dipoles
($ r\sim 1/k_{\perp}\sim 1/Q$) have a small ($\sim r^2$) total interaction
cross section with the proton.\\
b) Large size configurations with $r\sim 1/ \Lambda_{QCD}$ and
 $ k_{\perp}\sim \Lambda_{QCD}\ll Q$. They have a large total interaction
cross section, but contribute with a small phase space at large $Q^2$, because
 these configurations are kinematically possible only if the fraction of
 longitudinal momentum carried by one of the quarks is very small
 $x_1\sim k_{\perp}^2/Q^2\ll 1$. This configuration corresponds to the "aligned jet",
introduced by Bjorken and Kogut.\cite{BJ}

Both configurations lead to the same behaviour of $\sigma_{\gamma^*p}\sim 1/Q^2$,
but they behave differently in the process of the diffraction dissociation of
a virtual photon~\cite{Frank,Nik}. The cross section of such a process is
proportional to a square of modulus of the corresponding diffractive
amplitude and for a small size configuration it is small ($\sim 1/Q^4$). For
large size configurations a smallness is only due to the phase space
and the inclusive cross section for diffractive dissociation of a virtual
photon decreases as $1/Q^2$, i.e. in the same way as the total cross section.
This is true only for the total inclusive diffractive cross section, where
characteristic masses of produced states are $M^2\sim Q^2$. For exclusive
channels with fixed mass (for example production of vector mesons) situation
is different and these cross sections decrease faster than $1/Q^2$ at large
$Q^2$.

Inclusive diffractive production of very large masses ($M^2\gg Q^2$) can be described in
the first approximation by triple-Regge diagrams (Fig.8).\cite{15r} From the point of
view of the quark-gluon fluctuation of the fast photon triple-Pomeron contribution
corresponds to diffractive scattering of very slow (presumably gluonic) parton,
which has a small virtuality.

The model~\cite{CFKS} uses the picture of diffraction
dissociation of a virtual photon outlined above and is a natural
 generalization of models used for the description of high-energy hadronic
 interactions. The interaction of the small size component in the wave function
of a virtual photon is calculated using QCD perturbation theory. The main
 parameter of the model -
 intercept of the Pomeron was fixed from
 a phenomenological study of these interactions discussed above
 ($\Delta_P = 0.2$) and was found to give a good description of
 $\gamma^*p$-interactions in a broad range of $Q^2$ ($0\le Q^2< 10~GeV^2$).
 Another important parameter of the theory, the triple Pomeron vertex, obtained from
  a fit to the data ($r_{PPP}^{(0)}/g_{pp}^P(0) \approx 0.1$) is also in reasonable
 agreement with the analysis of soft hadronic interactions.\cite{Ponomar,15r}
 The description of $\sigma^{(tot)}_{\gamma^*p}$ as a function of s for different
 values of $Q^2$ ( experimental data are from H1~\cite{H1F2}, ZEUS~\cite{ZEUSF2})
  is shown in Fig.16.  Diffraction dissociation of a virtual
photon is usually presented as a function of $Q^2, M^2$ (or $\beta=Q^2/(M^2+Q^2)$)
and $x_P=x/\beta=(M^2+Q^2)/(W^2+Q^2)$. Description of HERA data on diffractive
dissociation~\cite{H1d} in the model is shown in Fig.17.
 The model reproduces experimental data quite well. It can be used to predict
 structure functions and partonic distributions at higher energies
or smaller $x$, which will be accessible for experiments at LHC.
\begin{figure}
\label{fig:sigmat}
\centering
\includegraphics[width=5.0in]{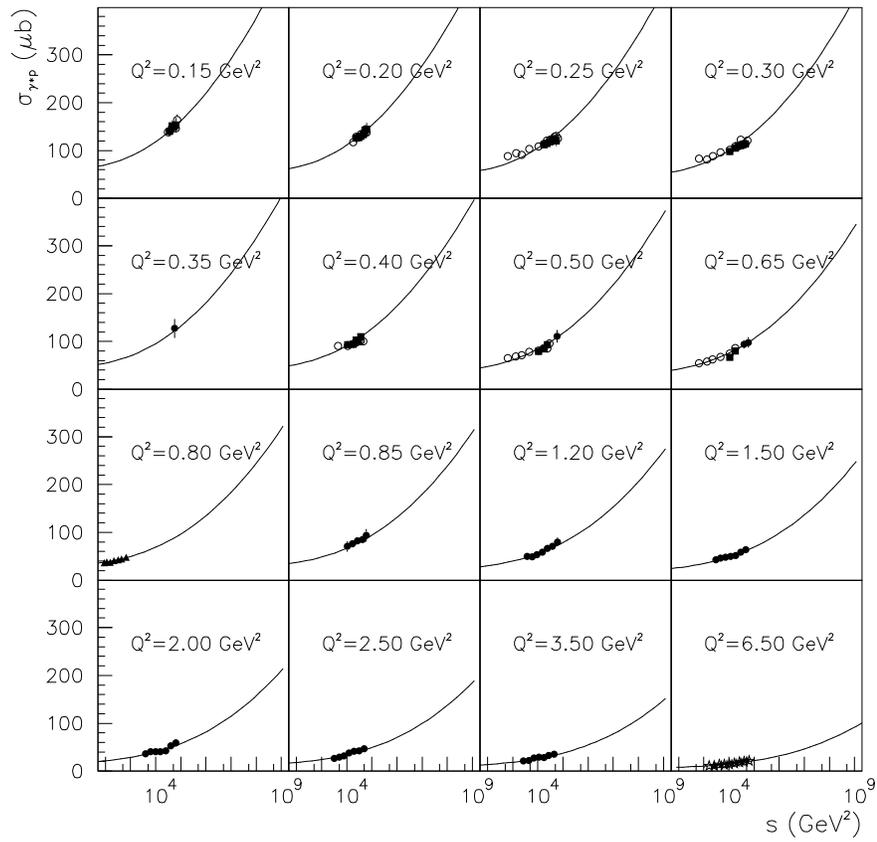}
\vskip 0.1in
\caption{$\sigma^{(tot)}_{\gamma^*p}$ as a function of s for different values of
 $Q^2$ compared with experimental data.}
\end{figure}
 \begin{figure}
\vskip0.05in
\centering
\includegraphics[width=5.0in]{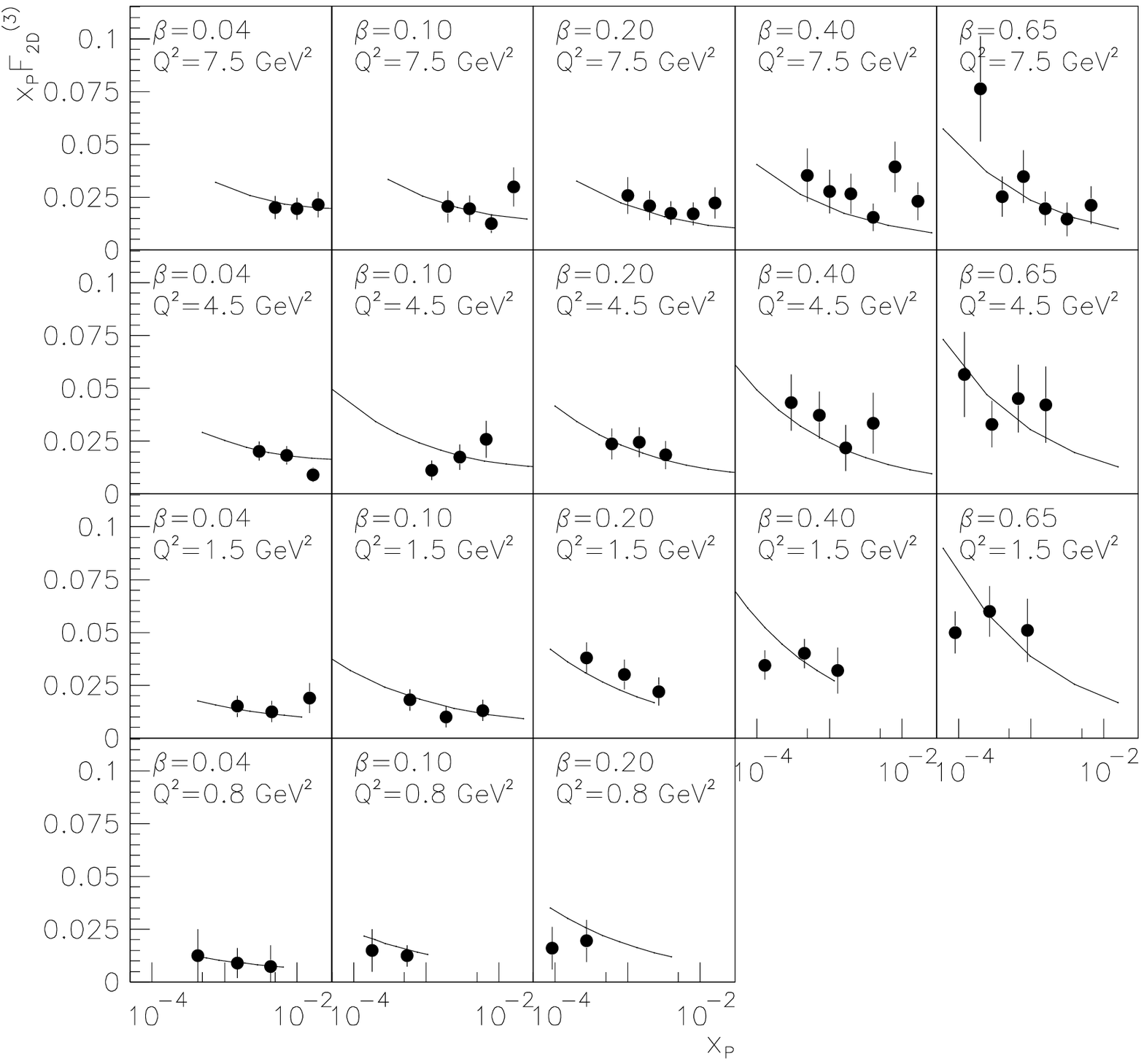}
\caption{ The diffractive structure function $x_P F_2^{D(3)}$ as a function
of $x_P$ for fixed values of $Q^2$ and $\beta=Q^2/(Q^2+M^2)$.
\label{fig:diff2}}
\end{figure}

 \section{Acknowledgments}

 I am grateful to K.Boreskov, A.Capella, V.Fadin, E.G. Ferreiro, O.V. Kancheli,
 J.H. Koch, G. Korchemski, E. Levin, L.N. Lipatov, C. Merino, C.A. Salgado,
 K.A. Ter-Martirosyan and J. Tran Thanh Van for useful discussions.\\
  This work is supported in part by the grants RFFI-98-02-17463 and NATO
  OUTR.LG 971390.

 \end{document}